\documentclass[conference, 9pt]{IEEEtran}
\IEEEoverridecommandlockouts
\usepackage{amsmath,amssymb,amsfonts}
\usepackage{algorithmic}
\usepackage{graphicx}
\usepackage{textcomp}
\usepackage{xcolor}
\usepackage{xspace}

\usepackage{times}  
\usepackage{helvet}  
\usepackage{courier}  
\usepackage[hyphens]{url}  
\usepackage{graphicx} 
\usepackage{caption} 

\usepackage{algorithm}
\usepackage{algorithmic}
\usepackage{amsmath}
\usepackage{booktabs}
\usepackage{tabularx}
\usepackage{array}
\usepackage{float}
\usepackage{multirow}
\usepackage{adjustbox}
\usepackage{afterpage}
\usepackage{placeins}
\usepackage{enumerate}
\usepackage{xcolor}
\usepackage[normalem]{ulem}
\usepackage{subcaption}
\usepackage{etoolbox}

\def\BibTeX{{\rm B\kern-.05em{\sc i\kern-.025em b}\kern-.08em
    T\kern-.1667em\lower.7ex\hbox{E}\kern-.125emX}}

\newcommand{\fvrulegen}{\textsc{FVRuleLearner}\xspace}
\newcommand{\nlsva}{NL-to-SVA\xspace}

\newcommand{\HumanDataset}{\textsc{NL2SVA-Human}\xspace}
\newcommand{\MachineDataset}{\textsc{NL2SVA-Machine}\xspace}
\newcommand{\IBMDataset}{\textsc{NL2SVA-OpenCore}\xspace}

\newcommand{\GPTFo}{\textsc{GPT-4o}\xspace}

\newcommand{\GPTotmini}{\textsc{o3-mini}\xspace}

\newcommand{\Claudefpf}{\textsc{Claude-4.5 Sonnet}\xspace}

\newcommand{\fvevaliclZ}{\textsc{FVEval-0-shot\xspace}}
\newcommand{\fvevaliclT}{\textsc{FVEval-3-shot\xspace}}
\newcommand{\emsemble}{\textsc{Ensemble\xspace}}

\newcommand{\BLEU}{\textsc{BLEU\xspace}}
\newcommand{\syntax}{\textsc{Syn\xspace}}
\newcommand{\func}{\textsc{Func\xspace}}
\newcommand{\rfunc}{\textsc{R.Func\xspace}}
\newcommand{\optree}{\textsc{Op-Tree\xspace}}

\newcommand{\Functionality}{\textsc{Functionality\xspace}}
\newcommand{\RelaxedFunc}{\textsc{Relaxed Functionality\xspace}}

\newcommand{\numData}{three\xspace}

\newcommand{\Jasper}{\textsc{Jasper}\xspace}

\renewcommand\footnoterule{%
  \kern -3pt
  \hrule width 0.4\columnwidth
  \kern 2.6pt
}

\begin{document}

\IEEEoverridecommandlockouts

\title{\fvrulegen: Operator-Level Reasoning Tree (\optree)-Based Rules Learning for Formal Verification\\
}


\author{
    \IEEEauthorblockN{
        Lily Jiaxin Wan$^{1,2,*}$, 
        Chia-Tung Ho$^{2,*}$, 
        Yunsheng Bai$^{2}$, 
        Cunxi Yu$^{2,3}$, 
        Ghaith Bany Hamad$^{2}$, 
        Deming Chen$^{1}$, 
        Haoxing Ren$^{2}$
    }
    \IEEEauthorblockA{$^{1}$University of Illinois Urbana-Champaign, $^{2}$NVIDIA, $^{3}$University of Maryland, College Park}
    \thanks{*Corresponding authors: Lily Jiaxin Wan (wan25@illinois.edu) and Chia-Tung Ho (chiatungh@nvidia.com).}
    \vspace{-5mm}
}

\setlength{\floatsep}{1cm}

\maketitle

\begin{abstract}
The remarkable reasoning and code generation capabilities of large language models (LLMs) have recently motivated increasing interest in automating formal verification (FV), a process that ensures hardware correctness through mathematically precise assertions but remains highly labor-intensive, particularly through the translation of natural language into SystemVerilog Assertions (\nlsva).
However, LLMs still struggle with \textcolor{black}{SVA generation} due to limited training data and the intrinsic complexity of FV operators. 
\textcolor{black}{Consequently, a more efficient and robust methodology for ensuring correct SVA operator selection is essential for producing functionally correct assertions.}
To address these challenges, we introduce \fvrulegen, an Operator-Level Rule (Op-Rule) learning framework built on a novel Operator Reasoning Tree (\optree), which models SVA generation as structured, interpretable reasoning.
\fvrulegen operates in two complementary phases: (1) Training: it constructs \optree\ that decomposes NL-to-SVA alignment into fine-grained, operator-aware questions, combining reasoning paths that lead to correct assertions; and (2) Testing: it performs operator-aligned retrieval to fetch relevant reasoning traces from the learned \optree\ and generate new rules for unseen specifications. 
In the comprehensive studies, the proposed \fvrulegen\ outperforms the
state-of-the-art baseline by 3.95\% in syntax correctness and by
31.17\% in functional correctness on average. 
Moreover, \fvrulegen\ successfully reduces an average of 70.33\% of SVA functional failures across diverse operator categories through a functional taxonomy analysis, showing the effectiveness of applying learned \optree\ to the Op-Rule generations for unseen NL-to-SVA tasks.
These results establish \fvrulegen as a new paradigm for domain-specific reasoning and rule learning in formal verification. 
The source code and benchmark are available at 
\textcolor{black}{https://github.com/NVlabs/FVRuleLearner}.
\end{abstract}

\begin{IEEEkeywords}
Large Language Models (LLMs), Formal Verification, SystemVerilog Assertions (SVAs)
\end{IEEEkeywords}
\vspace{-2mm}
\section{Introduction}

\begin{figure}
\centering
\includegraphics[width=0.366\textwidth]{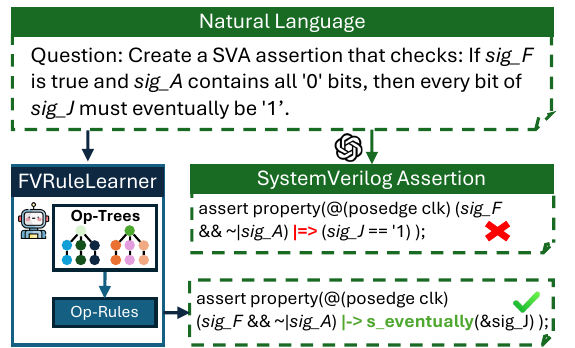}
\caption{\small{Given a \nlsva task, LLM inference often produces functionally incorrect SVAs, whereas our \fvrulegen\ framework leverages learned \optree\ to generate functionally correct SVAs.}}
\label{fig:nl2sva}
\vspace{-0.6cm}
\end{figure}

In the rapidly evolving hardware industry, the increasing complexity of integrated circuits (ICs) poses significant challenges to the verification process, a critical phase that accounts for nearly half of the total design cycle \cite{Wilson2021}.
Traditional simulation-based verification struggles to achieve comprehensive coverage, motivating the rise of formal verification (FV) as a complementary methodology that expresses design intent as temporal logic properties and \textcolor{black}{exhaustively} proves their correctness under all possible input conditions~\cite{hasan2015formal, seligman2023formal}.
However, writing SVAs from natural-language specifications (\nlsva) remains a major bottleneck: it is labor-intensive, error-prone, and requires substantial expertise in both hardware design and formal methods. Despite extensive literature and standardized assertion libraries \cite{vijayaraghavan2005practical,cerny2015sva,cummings2009systemverilog,foster2006introduction}, \textcolor{black}{the manual construction of assertions is still tedious and prone to mistakes, which has naturally motivated the development of automated assertion-generation methods, including recent LLM-based approaches. Unfortunately, current LLM systems still fall short in both accuracy and reliability for SVA generation. Fig.~\ref{fig:nl2sva} illustrates an example of these limitations.}



The challenges in translating NL specifications to SVAs (\nlsva) have sparked interest in applying Large Language Models (LLMs) to automate and improve productivity in digital design verification~\cite{liu2023chipnemo, kande2023llm, orenes2023using}. 
Despite their proficiency in generating code across various languages~\cite{nam2024using, lin2024llm}, their performance on formal verification tasks remains limited. 
Generated assertions often suffer from syntactic violations~\cite{chen2021evaluating, wang2024large} and, even when syntactically correct, frequently misrepresent the intended functional semantics~\cite{fang2024assertllm}, often failing to produce accurate SVA operators, especially for temporal implication and delay.
Consequently, a more efficient and robust methodology to ensure correct SVA operator selection is essential for generating functionally correct SVAs.

We propose~\fvrulegen, an operator-level rule learning framework that models assertion generation as a structured reasoning process over formal operators using novel \textbf{Op}erator Reasoning \textbf{Tree} (Op-Tree).  
\textcolor{black}{\fvrulegen\ constructs operator-level reasoning traces using the proposed \optree\ during training. At testing time, these reasoning traces are retrieved and transformed into Operator-level Rules (Op-Rules) through a rule-adaptation process that enables effective generalization for guiding SVA generation. Our contributions are summarized below:}

\begin{enumerate}
\item We developed an operator-level rule learning framework for SVA generation with the novel \optree, that models assertion generation as structured reasoning. These learned operator-level reasoning traces are retrieved to generate operator-level rules in the testing time, which enables LLMs to adapt and continuously improve on unseen examples without relying on expensive model fine-tuning.
\item We perform comprehensive studies to demonstrate the superior performance of \fvrulegen\ on functional correctness across the \numData datasets. \fvrulegen\ outperforms the state-of-the-art baseline by 3.95\% in syntax correctness and by 31.17\% in functional correctness on average.
\item We conduct a functional taxonomy analysis and demonstrate that \fvrulegen significantly reduces functional mismatches, achieving an average 70.33\% reduction in SVA functional failures across diverse operator categories relative to the state-of-the-art baseline.
\item We develop and release a new dataset of 1,000 \nlsva pairs from open-source hardware designs~\cite{opencores2024}, establishing a reproducible benchmark for LLM-assisted formal verification.
\end{enumerate} 

\section{Related Work}

\subsection{Formal Verification}

Early attempts at automated FV assertion generation using template-based or traditional ML methods faced limitations in handling the ambiguity and compositional structure of natural language specifications~\cite{bjorner1997automatic,hekmatpour2005block,yu2023survey}. 
Benchmarking efforts such as FVEval~\cite{FVEval2024} and AssertionBench~\cite{pulavarthi2024assertionbench} have established the first systematic evaluation protocols for LLMs in formal verification.
Building on these benchmarks, recent studies have explored LLMs for \nlsva translation, showing progress through document- or RTL-aware prompting and multi-stage reasoning pipelines, yet challenges remain in model generalization and explainability~\cite{kande2023llm,sun2023towards,orenes2023using,fang2024assertllm,mali2024chiraag,hassan2024llm,maddala2024laag}. 
The work by Orenes-Vera et al.~\cite{orenes2023using} manually designed rules to guide GPT-4 in generating better SVAs, whereas our approach focuses on learning from training data to automate this process.
More recently, Xiao et al.~\cite{Xiao2025HybridNL2SVA} combine retrieval-augmented generation (RAG) with task-specific fine-tuning to improve the alignment between natural language specifications and generated assertions.

\vspace{-0.025cm}

\subsection{LLM for Code and Hardware Design}

LLMs have shown notable proficiency in code generation across various programming languages~\cite{cen2024sqlfixagent,endres2024can,he2024beyond,mundler2024code,song2024effective,yin2024thinkrepair,ho2024large}, with advancements like Claude-Sonnet-4.5~\cite{anthropic_sonnet4_5} and DeepSeek-Coder-V2~\cite{deepseekcoderV2} leading the way. 
Recent efforts in Electronic Design Automation (EDA) have leveraged LLMs for tasks such as HDL generation and analysis~\cite{liu2023chipnemo,tsai2023rtlfixer, ho2025verilogcoder, Wan2025SchemaCoder}, yet the scarcity of high-quality, domain-specific datasets remains a barrier~\cite{he2023chateda}. 
However, their methodologies are limited and cannot be directly applied to the formal verification task~\cite{kande2023llm,sun2023towards,orenes2023using}.

\vspace{-0.025cm}

\subsection{LLM Self-Learning and Learning from Mistakes}
Studies have introduced LLM-based systems incorporating self-learning, self-correction, learning from previous mistakes through various approaches~\cite{liang2024internal,pan2023automatically,kamoi2024can}: supervised fine-tuning with reinforcement learning for code refinement~\cite{jiang2024training}, iterative refinement with self-feedback~\cite{madaan2024self}, simulators for self-correction in hardware design~\cite{huang2024towards}, linguistic feedback for trial-and-error learning~\cite{shinn2024reflexion}, Learning-From-Mistakes Prompting for low-resource language translation~\cite{liao2024learning}, recursive introspection~\cite{qu2024recursive}, retrieved in-context principles from previous mistakes~\cite{tong2024can,zhang2024context,sun2024retrieved}, pseudo-graph RAG for knowledge indexing and retrieval~\cite{liang2024empowering}, dual-agent system~\cite{yang2024collaborative}, agent and prompt optimization~\cite{yang2024large,zhangoffline,zhou2024symbolic,yuksekgonul2024textgrad}, etc. 
Unlike methods on model tuning or test case clustering, our approach first validates the effectiveness of learning operator-level reasoning traces during training and subsequently applies these Op-Rules to unseen cases.

\section{Preliminaries}
\label{sec:prelim}

\subsection{Problem Definition}
\label{subsec:problem-def}
Let $\mathcal{D} = \{(x_i, y_i)\}_{i=1}^N$ represent the dataset, where $x_i$ is a NL specification and $y_i$ is the corresponding golden SVA. 
Given an \nlsva task, we define the LLM inference as a function 
$F: \mathcal{X} \rightarrow \mathcal{Y}$, which maps NL specifications 
$x \in \mathcal{X}$ to predicted SVAs $\hat{y} = F(x) \in \mathcal{Y}$. The goal is to generate 
$\hat{y}$ that matches the golden SVA $y$, evaluated by metric $M$ measuring 
\textbf{syntax correctness}, \textbf{functionality}, and \textbf{linguistic similarity}. 
Here, \fvrulegen learns the operator-level reasoning traces to generate the Op-Rules $\mathcal{R}$ for maximizing metric $M$ for \nlsva tasks, formulated as
\begin{equation}
\mathcal{R}^\ast 
= \arg\max_{\mathcal{R}} 
\mathbb{E}_{(x,y) \sim \mathcal{D}} 
\left[ M\!\left(y, \hat{y}\right) \mid \mathcal{R} \right].
\label{Eq:Obj}
\end{equation}

\begin{table}[t]
\scriptsize
\centering
\caption{\small{Operator Categories and SystemVerilog Operators Table.}}
\label{tab:op-categories}
\renewcommand{\arraystretch}{1.15}
\begin{tabular}{p{0.23\textwidth} p{0.22\textwidth}}
\toprule[1.5pt]
\textbf{Operator Category} & \textbf{Representative Operators} \\
\midrule
Temporal Implication Operators & \texttt{|->}, \texttt{|=>} \\
Temporal Delay Operators & \texttt{\#\#m}, \texttt{[m:n]} \\
Temporal Sampling Operators & \texttt{\$past}, \texttt{\$rose}, \texttt{\$fell}, \texttt{\$stable} \\
Temporal Liveness Operators & \texttt{s\_eventually}, \texttt{s\_always}, \texttt{strong} \\
Combinational Logic Operators & \texttt{\&\&}, \texttt{||}, \texttt{!}, \texttt{==}, \texttt{!==}, \texttt{\^{}} \\
Miscellaneous Differences & \texttt{@(posedge clk)}, \texttt{disable iff(rst)} \\
\bottomrule[1.5pt]
\vspace{-0.6cm}
\end{tabular}
\end{table}

\begin{figure}
\centering
\vspace{-1cm}
\includegraphics[width=0.42\textwidth]{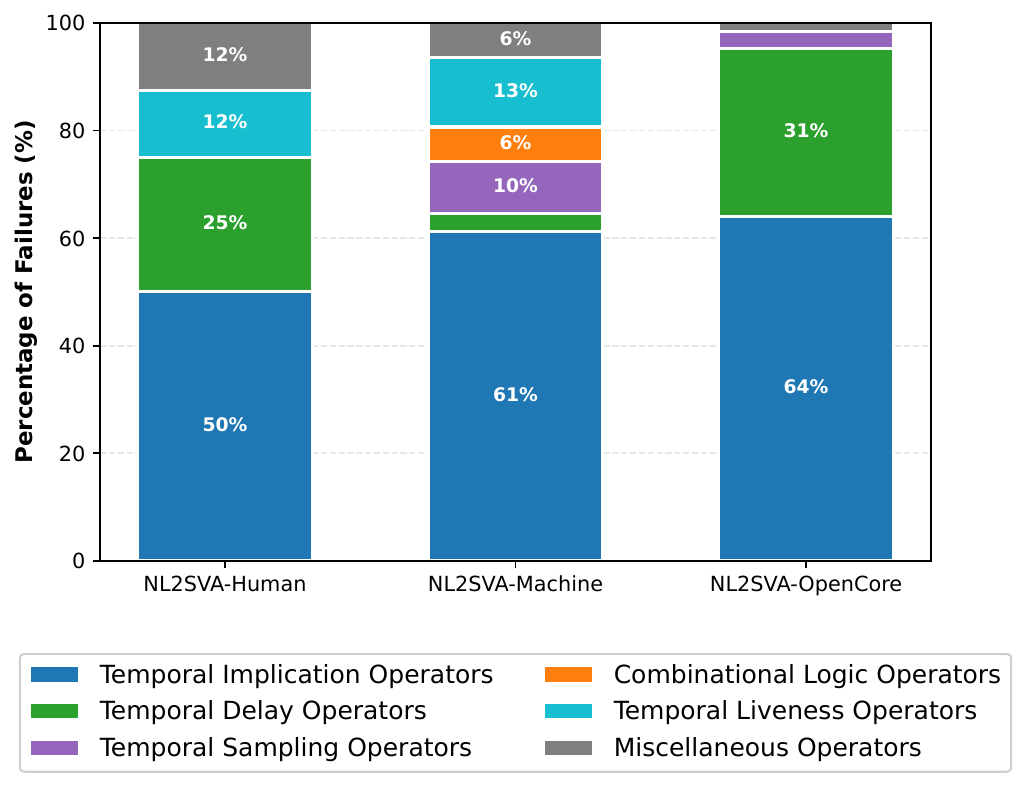}
\caption{\small{Operator-level functional mismatch analysis between generated and golden SVAs across three datasets. Temporal operators dominate the errors, highlighting the need for operator-aware reasoning.}}
\label{fig:operator}
\vspace{-0.5cm}
\end{figure}

\subsection{Operator-Level Functional Mismatch Analysis}
\label{subsec:operator-analysis}



We categorize SVA operators into six groups: temporal implication, temporal delay, temporal sampling, temporal liveness, combinational logic, and miscellaneous structural forms, as shown in Table~\ref{tab:op-categories}. We further analyze the functionally incorrect cases identified by the equivalence-checking tool to determine the operator-level failure modes.
We classify operator mismatches based on the generated SVA's operator rather than the golden SVA, explicitly highlighting the specific operators the model tends to hallucinate.

Fig.~\ref{fig:operator} shows that \textbf{\textit{temporal operators dominate functional mismatches in the \nlsva task}}, contributing to more than 80\% of all failures across datasets.
Among them, temporal implication operators (\texttt{|->}, \texttt{|=>}) alone cause 50–64\% of errors, showing that LLMs often confuse same-cycle and next-cycle semantics. 
Temporal delay and sampling operators introduce additional errors, indicating that LLMs struggle to represent timing intervals and signal transitions.
In contrast, combinational mismatches are rare, suggesting that LLMs handle Boolean logic more effectively. Overall, these findings indicate that the core difficulty in~\nlsva translation stems from insufficient temporal reasoning, highlighting the importance of operator-aware learning that captures event ordering and timing semantics.

\subsection{Operator-Level Rule (Op-Rule)}

\begin{figure}
\centering
\includegraphics[width=0.5\textwidth]{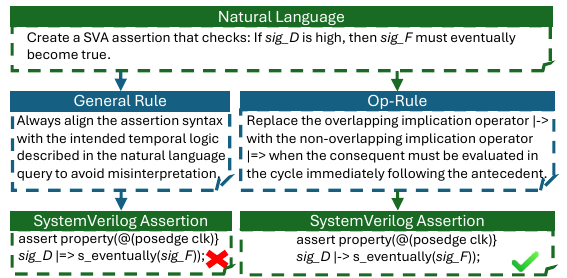}
\caption{\small{Comparison between General Rules and Op-Rules.}}
\label{fig:optree-general-comparison}
\vspace{-0.6cm}
\end{figure}

Unlike general rules which offer broad advice,
we tackle the challenges of using LLMs to understand operators, an Operator-Level Rule (Op-Rule) is a granular, executable directive targeting specific SVA constructs. 
We define an Op-Rule as a structured directive that explicitly maps a specific operator context to a precise syntactic modification.
As illustrated in Fig.~\ref{fig:optree-general-comparison}, the General Rule fails because it relies on abstract guidance to "align assertion syntax with intended temporal logic." 
Since LLMs struggle with the precise semantics of temporal operators, this ambiguity leads to incorrect SVA generation.
In contrast, the Op-Rule succeeds by enforcing execution over interpretation. By implementing a deterministic, symbol-level directive rather than relying on semantic reasoning, the Op-Rule ensures the generation of the correct overlapping implication.
\begin{figure*}[!t]
\centering
\includegraphics[width=0.92\textwidth]{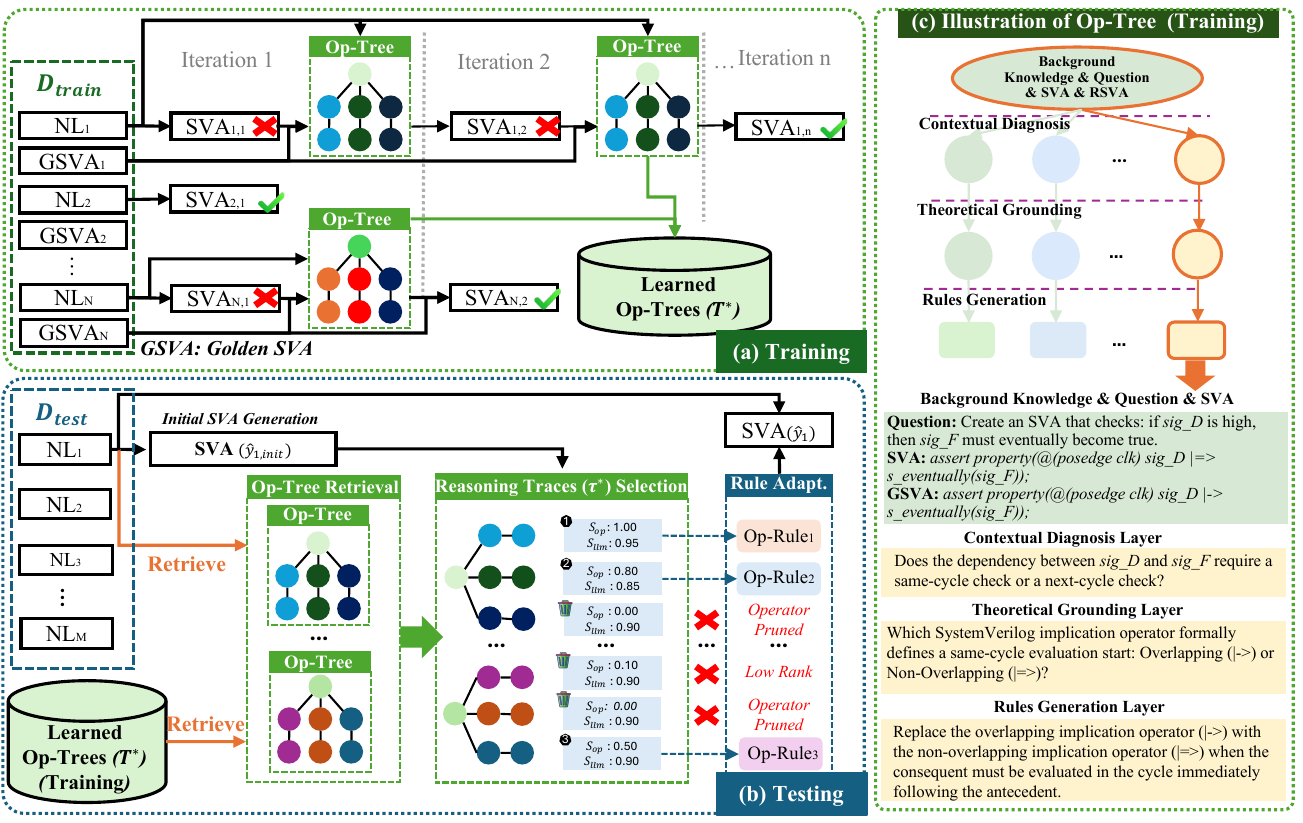}

\caption{
\small{The flow overview of \fvrulegen and an illustration of \optree~in training. (a) Training phase: The framework leverages \optree\ to consolidate the operator-level reasoning traces and they are stored in $\mathcal{T}^*$. (b) Testing phase: \fvrulegen\ firstly generates initial SVA ($\hat{y}_{init}$) and retrieve relevant \optree\ from $\mathcal{T}^*$ according to the similarity score of natural language specification of testing data and the background nodes' natural language content. Then, the candidate reasoning traces are selected based on the proposed operator alignment and semantic applicability scores. Lastly, the rule adaption step generate Op-Rules for final SVA ($\hat{y}$). (c) An \textbf{illustrative example} of the Op-Tree reasoning process, showing how hierarchical decomposition links design context to theoretical definitions to produce a verified correction rule.}}
\label{fig:method}
\vspace{-0.5cm}
\end{figure*}

\section{Proposed Approach: \fvrulegen}
\label{sec-method}



We propose \fvrulegen, a novel framework that shifts the generation paradigm from 
monolithic text translation to explicit operator-level reasoning, effectively bridging the gap between natural language and the strict timeline semantics of SystemVerilog.
As illustrated in Fig.~\ref{fig:method}, our approach decomposes this challenge into two phases: a \textbf{Training Phase} that acts as a self-reflective engine to distill high-fidelity reasoning traces (Op-Trees) from failure cases, 
and a \textbf{Testing Phase} that uses a hybrid retrieval mechanism to enforce structural alignment on unseen specifications.
As a result, the dataset is divided into: (1) \textbf{Training set}: $\mathcal{D}_{\text{train}} = \{(x_i, y_i)\}_{i=1}^{N_{\text{train}}}$, and (2) \textbf{Testing set}: $\mathcal{D}_{\text{test}} = \{(x_i, y_i)\}_{i=1}^{N_{\text{test}}}$, where $N_{\text{train}}$ and $N_{\text{test}}$ are the sizes of the training and testing sets, respectively.

\subsection{Training Phase}

Given the training set $\mathcal{D}_{\text{train}}$, the framework 
\textcolor{black}{utilizes the golden SVA as an oracle to diagnose mismatches and} invokes \textit{Op-Tree reasoning} to derive corrective operator-level steps. 
The validated traces form the verified Op-Tree set $\mathcal{T}$, which conditions 
the model through retrieved and regenerated operator rules for unseen inputs. The training objective is therefore to learn the optimal Op-Tree reasoning traces:
\begin{equation}
\small{
\begin{aligned}
&\hat{y} = F_{\mathcal{T}}(x) = F(x, r(x, y, \mathcal{T}))\\
&\mathcal{T}^\ast = \arg\max_{\mathcal{T}}
\mathbb{E}_{(x,y) \sim \mathcal{D}_{\text{train}}}
\left[ M\!\left(y, F_{\mathcal{T}}(x)\right) \right]
\end{aligned}
}
\label{Eq:TrainObj}
\end{equation}
Here, $r$ is the function to generate the Op-Rules $\mathcal{R}$ from the reasoning traces of \optree~($\mathcal{T}$) with the NL ($x$), golden SVA ($y$).
$F(x, r)$ represents the LLM inference with NL and the corresponding $\mathcal{R}$ as input and outputs the generated SVAs.

\noindent{\textbf{Operator Reasoning Tree (\optree):}}
To diagnose assertion failures, we introduce the Operator Reasoning Tree (\optree), a hierarchical reasoning structure generated via operator-directed prompting. 
Unlike linear chain-of-thought, the \optree~ mimics the systematic debugging workflow of a verification engineer, decomposing the error into three orthogonal layers (as shown in Fig.~\ref{fig:method}(c):

\begin{enumerate}
    \item \textbf{Contextual Diagnosis:} Analyzes specific signal relationships in the design to formulate a hypothesis about the failure mode.
    
    \item \textbf{Theoretical Grounding:} Validates this hypothesis by retrieving standard definitions (e.g., \textit{``Formal distinction between Overlapping `\texttt{|->}' and Non-Overlapping `\texttt{|=>}' implication''}), anchoring the reasoning to formal axioms.
    
    \item \textbf{Rule Generation:} Synthesizes the specific context and theoretical grounding into a correction Op-Rule at the leaf node.
\end{enumerate}
 
\textcolor{black}{To ensure broad reasoning coverage, we let LLMs focus on different aspects, to detect and discourage repetitive questions.}
This encourages the model to explore different error categories such as timing or logic rather than simply rephrasing the same hypothesis.

\noindent{\textbf{Training Iterations:}}
To guarantee the reliability of the learned rules, we enforce a strict acceptance criterion. As shown in Iteration 2 of Fig.~\ref{fig:method}, the rules extracted from the \optree~ are fed back into the system to regenerate the assertion ($\text{SVA}_{1,2}$). One \optree~is deemed ``Valid'' and committed to the database if and only if this rule-guided iteration produces an SVA that is functionally equivalent to the golden reference (marked by the green check $\checkmark$). This self-reflection process effectively ensures the system learns exclusively from empirically verified correction strategies.

\subsection{Testing Phase}



The testing phase executes a dynamic, retrieval-augmented inference pipeline to adapt learned reasoning to unseen specifications. 
For a given input $x \in \mathcal{D}_{\text{test}}$, we firstly generate an initial SVA $\hat{y}_{init}$ without Op-Rules. 
We then retrieve the relevant learned \optree\ $\Bar{T}$ from $\mathcal{T}^*$ by selecting the top-k similarity score of the $x$, and the background node of each learned \optree.
The reasoning traces $\boldsymbol{\tau}$ are extracted from root node to leaf nodes of each \optree\ in $\bar{T}$. 
Then, the top-k reasoning traces $\boldsymbol{\tau}^*$ are selected using a novel hybrid score of operator and semantic by referencing $\hat{y}_{init}$ and $x$ for generating Op-Rules ${\mathcal{R}}^*$ in rule adaption phase ($ra(x, \boldsymbol{\tau}^*)$) for generating the final SVA for $x$ in Eq.~\ref{Eq:Inference}.

\begin{equation}
\small{
\hat{y} = F(x, {\mathcal{R}}^*) = F(x, ra(x, \boldsymbol{\tau}^*)), \boldsymbol{\tau}^* \in \bar{T}
}
\label{Eq:Inference}
\end{equation}
Here, $ra$ denotes the rule adaptation procedure that transforms the selected reasoning traces into the Op-Rules ${\mathcal{R}}^*$.
\textcolor{black}{We empirically use $k=3$.}
We will introduce the novel hybrid score that considering operator and semantic for reasoning traces selection and rule adaption below.





\noindent \textbf{Hybrid Score for Reasoning Traces Selection}: 
Standard RAG systems rely on linguistic similarity, which is often a poor proxy for verification logic (e.g., keywords match but semantics differ).
To bridge this gap, we evaluate every candidate reasoning trace $\boldsymbol{\tau}$ considering operator and semantic as Eq.~\ref{eq4}.

\begin{equation}
\label{eq4}
\small{
S_{hybrid} = 
\begin{cases} 
    0, & \text{if } S_{op} = 0 \text{ or } S_{llm} = 0 \\
    \alpha \cdot S_{op} + (1-\alpha) \cdot S_{llm}, & \text{otherwise}
\end{cases}
}
\end{equation}

where $\alpha$ is a hyperparameter balancing operator and semantic weights.
\textcolor{black}{We empirically set $\alpha=0.5$ for the experiments.} The components are defined as follows:

\begin{itemize}
    \item \textbf{Operator Alignment Score ($S_{op}$):} This metric quantifies operator relevance between the candidate reasoning traces and $\hat{y}_{init}$.
    Let $\mathbb{O}$ denote the universe of valid operators defined in Table I. 
    We define the operator extraction function $\Phi$ as the projection of each reasoning trace ${\tau}^*$ and $\hat{y}_{init}$ in text format onto a operator subset $\mathcal{O} \in \mathbb{O}$.
    The extracted operator subset of $\hat{y}_{init}$, and $\boldsymbol{\tau}^*$ are $\mathcal{O}_{init} = \Phi(\hat{y}_{init})$, and $\mathcal{O}_{trace} = \Phi({\tau}^*)$, respectively.
    To robustly handle operator variations (e.g., matching $|{\rightarrow}$ with $|{\Rightarrow}$), we compute the alignment using a \textit{soft Jaccard Similarity}:
    \begin{equation}
    \label{eq5}
        S_{op}(\tau^*) = \frac{\sum_{u \in \mathcal{O}_{init}} \max_{v \in \mathcal{O}_{trace}} \text{sim}(u, v)}{|\mathcal{O}_{init} \cup \mathcal{O}_{trace}|}
    \end{equation}
    \textcolor{black}{Here, $\text{sim}(u,v)\in[0,1]$ assigns $1$ to identical operators, 0.5 to same category but not identical one (Table~\ref{tab:op-categories}), and $0$ otherwise.}
    
    \item \textbf{Semantic Applicability Score ($S_{llm}$):} This metric ensures reasoning quality. We employ an LLM-as-a-judge to evaluate the reasoning chain's transferability~\cite{zheng2023judging}. 
    We employ an LLM-as-a-judge~\cite{zheng2023judging} to assign a confidence score $S_{llm} \in [0,1]$ for each reasoning trace ${\tau}^*$ and natural language specification $x$.
\end{itemize}

The final ranking ensures that selected reasoning traces $\boldsymbol{\tau}^*$ are both linguistically relevant and operator aligned.

\noindent \textbf{Rule Adaption}:
While retrieved reasoning traces $\boldsymbol{\tau}^*$ capture high-fidelity reasoning patterns, they remain semantically coupled to the specific signal names or parameters (i.e., clock cycle) from training dataset $\mathcal{D}_{\text{train}}$.
To enable generalization at testing time, we apply a signal abstraction mechanism that masks instance-specific variables with generic placeholders (e.g., \texttt{<signal>}), enabling the model to focus on operator-level reasoning.
This step injects formal definitions directly into the context window, enforcing strict adherence to SystemVerilog semantics.
Guided by this context, the LLM generates new Op-Rules $\mathcal{R}^*$ adapted to the target specification to generate the final SVA as mentioned in Eq.~\ref{Eq:Inference}. 
\section{Experiments}

\begin{table*}[h!]
\centering
\small      
\caption{\small{Experimental Results of LLM Models on \HumanDataset, \MachineDataset, and \IBMDataset\ Testing Sets.}} 
\label{tab:llm-results}
{%
\begin{tabular}{l*{16}{c}}
\toprule[1.5pt]
\multirow{3}{*}{\textbf{Model / Method}} & 
\multicolumn{4}{c}{\textbf{\HumanDataset}} & 
\multicolumn{4}{c}{\textbf{\MachineDataset}} & 
\multicolumn{4}{c}{\textbf{\IBMDataset}} \\
\cmidrule(lr){2-5} \cmidrule(lr){6-9} \cmidrule(lr){10-13}
& \BLEU & \syntax & \func & \rfunc 
& \BLEU & \syntax & \func & \rfunc 
& \BLEU & \syntax & \func & \rfunc \\
\midrule[1.5pt]

\multicolumn{13}{l}{\textbf{GPT-4o}} \\
\midrule
\fvevaliclZ & 0.5203 & 0.9333 & 0.5333 & 0.6667 & 0.6054 & 0.9500 & 0.4167 & 0.4500 & 0.6145 & 0.9250 & 0.7050 & 0.7200 \\
\fvevaliclT & 0.5823 & 0.9333 & 0.4000 & 0.6000 & 0.6461 & 0.9500 & 0.4333 & 0.5167 & 0.6187 & 0.8950 & 0.6750 & 0.6950 \\
\emsemble   & 0.5876 & 0.8667 & 0.4000 & 0.4667 & 0.6562 & 0.9667 & 0.5000 & 0.6000 & 0.5922 & 0.9550 & 0.7000 & 0.7150 \\
\fvrulegen  & \textbf{0.6846} & \textbf{1.0000} & \textbf{0.8000} & \textbf{0.8000} & \textbf{0.7072} & \textbf{0.9667} & \textbf{0.7500} & \textbf{0.8167} & \textbf{0.6976} & \textbf{0.9900} & \textbf{0.8500} & \textbf{0.8750} \\

\midrule[1.5pt]
\multicolumn{13}{l}{\textbf{Claude 4.5 Sonnet}} \\
\midrule
\fvevaliclZ & 0.5330 & \textbf{1.0000} & 0.6000 & \textbf{0.8667} & 0.6246 & 0.9167 & 0.5333 & 0.6167 & 0.5868 & 0.9350 & 0.6900 & 0.7000 \\
\fvevaliclT & 0.6480 & \textbf{1.0000} & 0.5000 & 0.5000 & 0.6765 & 0.9500 & 0.4667 & 0.5667 & 0.6154 & 0.9150 & 0.7050 & 0.7250 \\
\emsemble   & 0.5324 & \textbf{1.0000} & 0.5333 & 0.8000 & 0.6212 & 0.9500 & 0.5333 & 0.6333 & 0.5881 & 0.9600 & 0.7050 & 0.7250 \\
\fvrulegen    & \textbf{0.7416} & \textbf{1.0000} & \textbf{0.8667} & \textbf{0.8667} & \textbf{0.7038} & \textbf{0.9667} & \textbf{0.7500} & \textbf{0.8167} & 
\textbf{0.6901} & \textbf{0.9700} & \textbf{0.8650} & \textbf{0.8900} \\

\midrule[1.5pt]
\multicolumn{13}{l}{\textbf{o3-mini}} \\
\midrule
\fvevaliclZ & 0.5886 & 0.9333 & 0.4667 & 0.6000 & 0.6075 & 0.9333 & 0.4500 & 0.5333 & 0.5900 & 0.9600 & 0.6800 & 0.7000 \\
\fvevaliclT & 0.6360 & 0.9333 & 0.4667 & 0.5000 & 0.6696 & 0.9500 & 0.4833 & 0.5833 & 0.6189 & 0.8950 & 0.6800 & 0.7150 \\
\emsemble & 0.6127 & 0.9333 & 0.4667 & 0.5500 & 0.6127 & 0.9333 & 0.4667 & 0.5500 & 0.5900 & 0.9600 & 0.6800 & 0.7000 \\
\fvrulegen   & \textbf{0.7363} & \textbf{1.0000} & \textbf{0.8667} & \textbf{0.9333} & \textbf{0.6980} & \textbf{0.9667} & \textbf{0.7833} & \textbf{0.8333} & \textbf{0.7145}	 & \textbf{0.9950}	 & \textbf{0.9150}	 & \textbf{0.9300} \\

\bottomrule[1.5pt]
\multicolumn{13}{l}{\small Note: \BLEU = BLEU score, \syntax = Syntax, \func = Functionality, \rfunc = Relaxed Functionality.} \\
\end{tabular}%
\vspace{-0.15cm}
}
\end{table*}



In this section, we present the comprehensive experimental results of applying our framework to various Large Language Models (LLMs) and compare their performance against state-of-the-art results from existing literature.

\subsection{Experimental Settings}

\noindent {\bf Dataset:} Our experiments were conducted using \numData distinct datasets:
\HumanDataset\ and \MachineDataset\ \cite{FVEval2024} are specifically curated for natural-language to SVA generation tasks. \textcolor{black}{In addition, we include \IBMDataset, a new dataset derived from \textsc{AssertionBench}~\cite{pulavarthi2024assertionbench}, consisting of 1,000 NL--SVA pairs mined from real-world OpenCore designs, targeting SVA generation for module interfaces.}
Each dataset is randomly split into 80\% training and 20\% testing subsets to ensure rigorous comparison and evaluation of the models' performance.

\noindent {\bf Evaluation Metrics:} We use four key metrics: 
(1) \textit{\BLEU}: Measures the textual similarity between generated and reference assertions, ranging from 0 (no similarity) to 1 (perfect match); 
(2) \textit{\syntax}: Evaluates the syntactic correctness of generated assertions (0 or 1);
(3) \textit{\Functionality}: Strictly evaluates whether the generated assertion's behavior precisely matches the reference (0 or 1);
(4) \textit{\RelaxedFunc}: Evaluates functional equivalence, allowing for implication relationships between the generated and reference assertions (0 or 1). An implication relationship here means if the reference assertion holds, the generated assertion must also hold, allowing for more permissive generated assertions. 
\Jasper~\cite{jasper2024} is further used as both the syntax checker during the testing phase and as an evaluation tool for assessing assertion functional correctness with two sets of scripts.

\begin{figure*}[!t] 
    \centering
    
    \begin{subfigure}[b]{0.32\textwidth} 
        \centering
        \includegraphics[width=\linewidth]{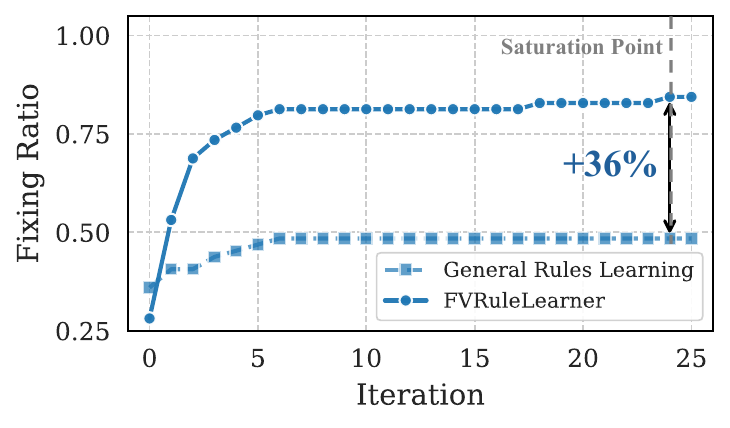}
        \caption{\HumanDataset}
        \label{fig:trend_human}
    \end{subfigure}
    \hfill 
    \begin{subfigure}[b]{0.32\textwidth}
        \centering
        \includegraphics[width=\linewidth]{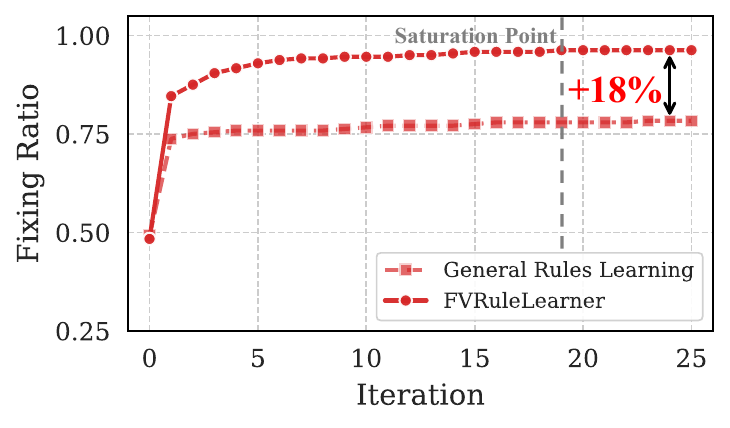}
        \caption{\MachineDataset}
        \label{fig:trend_machine}
    \end{subfigure}
    \hfill 
    \begin{subfigure}[b]{0.32\textwidth}
        \centering
        \includegraphics[width=\linewidth]{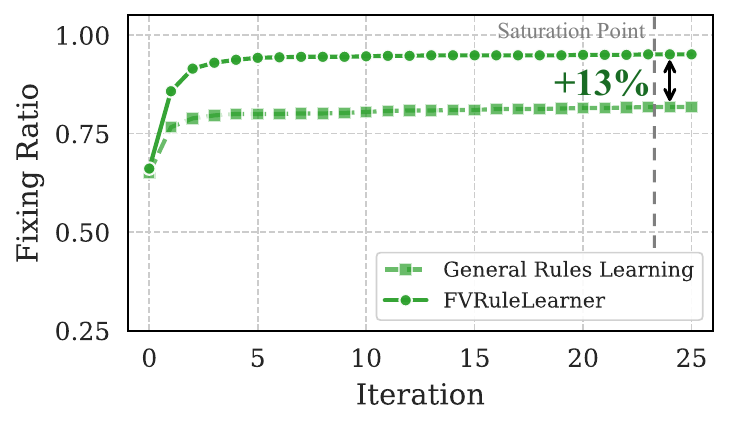}
        \caption{\IBMDataset}
        \label{fig:trend_opencore}
    \end{subfigure}

    \caption{
    \small{Comparison of fixing ratio trends over training iterations for \fvrulegen (circle) versus General Rule Learning (square) methods. Across (a) \HumanDataset, (b) \MachineDataset, and (c) \IBMDataset, the proposed \optree\ rules learning methodology consistently accelerates convergence and achieves higher final accuracy, improving the fixing rate by an average of 23.33\%.}}
    \label{fig:three_trends}
\vspace{-0.4cm}
\end{figure*}

\noindent {\bf Baselines:} 
\textcolor{black}{We compare \fvrulegen\ against several state-of-the-art baselines with publicly available implementations, including the zero-shot (\fvevaliclZ) and few-shot (\fvevaliclT) prompting configurations from FVEval~\cite{FVEval2024}.}
Additionally, we evaluate an~\emsemble~\cite{niimi2025simple}, which aggregates multiple inference passes from the base model using majority voting to improve robustness. 

Since each baseline, as well as our \fvrulegen framework, requires an LLM engine for inference, we consider a diverse selection of both general instruction-tuned models and reasoning models from two major commercial LLM families: (1)~\GPTFo \cite{openai2023gpt4},  (2)~\Claudefpf~\cite{anthropic_sonnet4_5},
(3)~\GPTotmini~\cite{openai_o3_mini}.

\subsection{Training Phase}

We show the functionality fixing ratio of the training set for each dataset over 25 iterations of the \fvrulegen against a general rule learning baseline in Fig.~\ref{fig:three_trends}.
The general rule learning baseline iteratively explore general rules for each \nlsva~problem, without \optree~reasoning.
\fvrulegen\ demonstrates superior convergence across all benchmarks, achieving an average 23.33\% improvement in fixing rates. 
\textcolor{black}{Moreover, \fvrulegen~achieves up to 36\% functional fixing rate on \HumanDataset.}
The steady performance gains across diverse specifications highlight that fine-grained operator-level reasoning effectively resolves the challenging SVA generation tasks. 
The remaining gap to 100\% corresponds to failures arising from highly coupled global dependencies or ambiguous intent, particularly within the NL descriptions in \HumanDataset.


\noindent{\textbf{Training Ratio Study:}} 
\textcolor{black}{Here, we analyze the \MachineDataset~to examine the functionality score trends under varying training data ratios, leveraging its larger scale to obtain statistically meaningful insights on testing performance.}
Here, we implemented an operator-level sampling strategy instead of using standard random sampling, since standard random sampling fails to capture this efficiency due to the \textit{long-tail distribution} of industrial specifications, where critical operators (e.g., temporal liveness, sampling operators) are statistically sparse compared to trivial combinational logic. 
We firstly categorize the data points in the training dataset based on Table~\ref{tab:op-categories}. 
Then, we evenly sample the data points from each categories, and remove the duplicated data points since a data point can be in different groups since its golden SVA contains operators in different categories.

\begin{figure}
\centering
\includegraphics[width=0.4\textwidth]{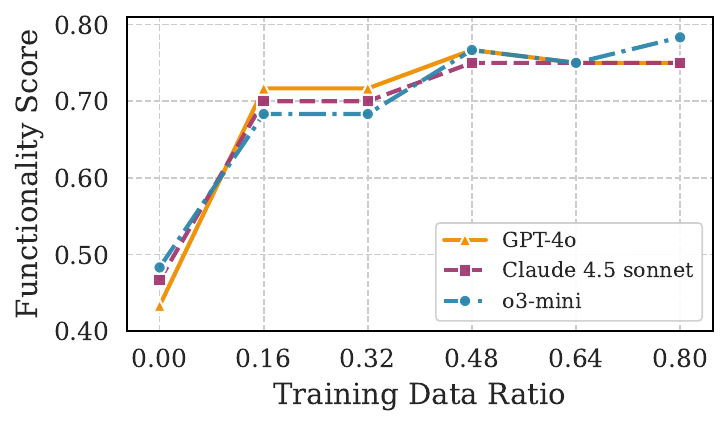}
\caption{
\small{\textcolor{black}{Effect of Training Data Ratio on \fvrulegen\ Performance on the \MachineDataset\ dataset.}}}
\label{fig:trainingratio}
\vspace{-0.7cm}
\end{figure}

The functional correctness improves sharply, saturating at a mere 48\% training ratio as shown in Fig.~\ref{fig:trainingratio}.  
From the study, we can observed that \fvrulegen performance depends on the diversity of Op-rules on operators from training stage rather than dataset scale.
In summary, these results show that \fvrulegen is highly data-efficient, effectively generalizing from a compact, high-density set of reasoning traces without the need for large-scale data overhead for training.
The maximum training data ratio is 80\%.


\subsection{Testing Phase}
\textcolor{black}{We show the main experimental results in Table~\ref{tab:llm-results} \textcolor{black}{that demonstrates} the efficacy of the \fvrulegen~framework across \numData diverse datasets and various state-of-the-art LLMs.}
On average, our~\fvrulegen~framework achieves \textcolor{black}{state-of-the-art 85.50\% functional correctness} across three datasets using~\GPTotmini, representing up to a 31.17\% improvement over the strongest existing baselines. \textcolor{black}{For~\GPTFo and \Claudefpf, \fvrulegen~further improves the average functionality score by 22.06\% and 18.11\%, respectively.}
In addition, \fvrulegen~achieves 98.39\% syntax correctness and 86.24\% relaxed functionality on average across the three datasets and all models.
The results consistently demonstrate improvements across all evaluated metrics, underscoring the framework’s robustness and adaptability.
Collectively, these results demonstrate that \fvrulegen~consistently produces high-quality assertions that are both syntactically correct and functionally accurate across various LLM models.



\noindent{\textbf{Rule Retrieval Ablation Study}}: We dissect the inference pipeline to distinguish the impact of rule \textit{granularity} versus \textit{adaptability}. We benchmark \fvrulegen~ against two variants: (1) \textit{General Rules}, which relies on broad rules generated during training; and (2) \textit{Static Op-Rules}, which applies Op-Rules from training directly without rule adaption for the testing data. 
\textcolor{black}{
Adaptation is crucial because static rules often reference generic signals that are absent in the specific design context; for example, a static rule might include \texttt{tb\_gnt}, which does not exist in the target design, leading to undefined signal errors if applied directly.
}
Table~\ref{tab:ablation-study-nl2sva-machine} shows that operator-level specificity offers a baseline improvement over general rules, and rule adaption process further boosts the performance, pushing functional accuracy to 78.33\%. 

\begin{table}[!tbp]
\centering
\scriptsize      
\caption{\small{Rule Retrieval Ablation Study Results for~\GPTotmini on \MachineDataset Dataset.
}}
\label{tab:ablation-study-nl2sva-machine}
{%
\begin{tabular}{lccc}
\toprule[1.5pt]
\textbf{Model / Method} & \textbf{\syntax} & \textbf{\func} & \textbf{\rfunc} \\
\multicolumn{4}{l}{\textbf{\GPTotmini}} \\
\midrule
\fvevaliclT & 0.9500 & 0.4833 & 0.5833\\
General Rules & 0.9500 & 0.6667 & 0.7500 \\
Static Op-Rules~ & 0.9500 & 0.7167 & 0.7833 \\
\fvrulegen & \textbf{0.9667} & \textbf{0.7833} & \textbf{0.8333} \\
\bottomrule[1.5pt]
\vspace{-0.8cm}
\end{tabular}%
}
\end{table}

\subsection{Runtime and Token Cost Analysis} 
Here, we report the runtime and token cost of \GPTotmini~for training, and testing.
Per \nlsva~pair, for the \HumanDataset~dataset, training the \optree~takes approximately 65.36 seconds and consumes 47.6k tokens with the approximate cost of \$0.12 per case; testing one assertion takes 3.92 seconds and 11.2k tokens with the cost of \$0.03. 
The \MachineDataset~dataset requires 82.68 seconds (9.0k tokens, \$0.02) for training and 48.86 seconds (9.0k tokens, \$0.02) for generation. 
Finally, the \IBMDataset~dataset requires 20.13 seconds (4.6k tokens, \$0.01) for training and 50.80 seconds (7.9k tokens, \$0.02) for generation. 
Both training and generation processes remain fully parallelizable. 
This shows the training procedure of \fvrulegen~is cost efficient and can improve the testing performance significantly.

\subsection{Functional Taxonomy Analysis at the Operator Level}
We analyze the efficiency of \fvrulegen on correcting functional failures of different operator categories (i.e., Section~\ref{subsec:operator-analysis}) compared to \fvevaliclT. In Table~\ref{tab:operator-failures}, \fvrulegen achieves a massive reduction in functional mismatches, decreasing total failures by an average of \textbf{70.3\%} across all datasets.
Crucially, the improvement is most pronounced in the categories where LLMs typically struggle the most:
\begin{itemize}
    \item \textbf{Temporal Implication Operators (80.6\% Reduction):}
The sharp decline in implication errors (e.g., swapping \texttt{|->} with \texttt{|=>}) indicates that the Op-Rules successfully guide the model to distinguish between \textit{overlapping} and \textit{non-overlapping} validity, which is a common semantic blind spot for standard LLMs.
    \item \textbf{Temporal Delay Operators (85.0\% Reduction):} The near-elimination of delay errors demonstrates that the retrieved rules provide precise guidance on cycle-accurate sequencing, correcting off-by-one errors that plague zero-shot generation.
\end{itemize}
This targeted reduction proves that \fvrulegen does not simply memorize patterns, but actively aligns the LLM's generation with the rigorous temporal semantics of SystemVerilog, effectively "debugging" the model's logic before the assertion is even finalized.




\begin{table}[t]
\centering
\scriptsize     
\setlength{\tabcolsep}{2pt} 
\caption{Operator-Level Failure Analysis showing the reduction in functional mismatches. The ``Red.'' column indicates the percentage reduction, where bold blue values highlight significant improvements achieved by \fvrulegen. ``Red.'' is the percentage reduction (Temp.=Temporal, Comb.=Combinational, Impl. =Implication, Sampl.=Sampling, Live.=Liveness).}
\label{tab:operator-failures}
\resizebox{\columnwidth}{!}
{%
\begin{tabular}{@{}lccccccccc@{}}
\toprule[1.5pt]
\multirow{2}{*}{\textbf{Op. Category}} & 
\multicolumn{3}{c}{\textbf{NL2SVA-Human}} & 
\multicolumn{3}{c}{\textbf{NL2SVA-Machine}} & 
\multicolumn{3}{c}{\textbf{NL2SVA-OpenCore}} \\
\cmidrule(lr){2-4} \cmidrule(lr){5-7} \cmidrule(lr){8-10}
 & Base & \textbf{Ours} & Red. & Base & \textbf{Ours} & Red. & Base & \textbf{Ours} & Red. \\
\midrule
Temp. Impl. & 4 & \textbf{1} & \textbf{\textcolor{blue}{-75\%}} & 19 & \textbf{4} & \textbf{\textcolor{blue}{-79\%}} & 41 & \textbf{5} & \textbf{\textcolor{blue}{-88\%}} \\
Temp. Delay       & 2 & \textbf{0} & \textbf{\textcolor{blue}{-100\%}} & 1 & \textbf{0} & \textbf{\textcolor{blue}{-100\%}} & 20 & \textbf{9} & \textbf{\textcolor{blue}{-55\%}} \\
Temp. Sampl.    & 0 & 0 & -- & 3 & \textbf{2} & \textbf{\textcolor{blue}{-33\%}} & 2 & \textbf{1} & \textbf{\textcolor{blue}{-50\%}} \\
Temp. Live.    & 1 & \textbf{0} & \textbf{\textcolor{blue}{-100\%}} & 4 & \textbf{3} & \textbf{\textcolor{blue}{-25\%}} & 0 & 0 & -- \\
Comb. Logic  & 0 & 0 & -- & 2 & 2 & 0\% & 0 & 0 & -- \\
Miscellaneous        & 1 & 1 & 0\% & 2 & \textbf{1} & \textbf{\textcolor{blue}{-50\%}} & 1 & 1 & 0\% \\
\midrule
\textbf{Total Failures} & 8 & \textbf{2} & \textbf{\textcolor{blue}{-75\%}} & 31 & \textbf{12} & \textbf{\textcolor{blue}{-61\%}} & 64 & \textbf{16} & \textbf{\textcolor{blue}{-75\%}} \\
\bottomrule[1.5pt]
\end{tabular}%
\vspace{-1cm}
}
\end{table}

\section{Conclusion and Future Work}
In this work, we address the semantic gap in LLM-based formal verification by introducing \fvrulegen, a framework shifting from text imitation to structured, operator-level reasoning. Our experiments identify an operator-level rule saturation phenomenon, \textcolor{black}{where functionality score plateaus once core reasoning rules are learned, regardless of further data scaling.}
\fvrulegen outperforms state-of-the-art methods by an average of 31.17\% in functional correctness and reduces functional failures by an average of 70.33\%, all while maintaining 98.56\% syntactic correctness. Additionally, we release \IBMDataset~to establish an industrial benchmark for future research.
\textcolor{black}{
Limited by model reasoning, future work will integrate \fvrulegen~into industrial pipelines and explore post-training (e.g., domain-specific fine-tuning).
}
 
\section*{Acknowledgement}

This work is supported by NSF 2117997 grant through the A3D3 institute, and Semiconductor Research Corporation (SRC) 2023-CT-3175 grant. 
We thank Ghaith Bany Hamad for his invaluable and profound technical insights and Syed Suhaib for his support.

\bibliographystyle{IEEEtran}
\bibliography{ref}

\end{document}